\documentclass{hep99}
\usepackage{epsfig}
\begin{document}

\title{Searches for Exotics at HERA}

\author{E. Perez$^1$\\ On behalf of the H1 and ZEUS Collaborations.}
%

\address{$^1$ CE-Saclay, DSM/DAPNIA/Spp\\[3pt]
E-mail: {\tt eperez@hep.saclay.cea.fr} \\
 Presented at the EPS'99 Conference, Tampere, Finland, 15-21 July 1999. }

\abstract{
Searches for contact interactions, leptoquark bosons and
excited fermions carried out at the HERA $ep$ collider are presented here.
The searches are based on $\sim 40 {\mbox{pb}}^{-1}$ of $e^+ p$ data
per experiment collected at a centre of mass energy
$\sim 300$ GeV.
First results on $e^- p$ data collected in 1998-1999 are
also presented.
}

\maketitle


\begin{picture}(80,100)
\put(350,310){DAPNIA/SPP 99-30}
\end{picture}
\vspace{-4.5cm}


\section{Contact Interactions}
Contact Interactions (CI) can be used to parameterize any new
physics process appearing at an energy scale $\Lambda$ above the
centre of mass energy $\sqrt{s}$. 
At HERA, $eeqq$ four-fermions terms would interfere 
(constructively or destructively) with
the Standard Model (SM) Neutral Current (NC) Deep Inelastic Scattering (DIS),
where the incoming lepton interacts with a quark coming from 
the proton, via the $t$-channel exchange of a virtual
gauge boson, such that the distributions of the Lorentz invariant
variables $Q^2$, $x$ and $y$ (related via $Q^2 = x y s$)
would be affected.

The H1~\cite{H1CI} and ZEUS~\cite{ZEUSCI} Collaborations 
searched for such distorsions in the full $e^+ p$ data collected
between 1994 and 1997,
using respectively a $\chi^2$ fit of the single differential cross-section
$d \sigma / d Q^2$, and a 2-dimensional likelihood analysis in $(x,y)$.
Various models (characterizing the structure of the CI) have been
constrained and the resulting lower bounds on 
the scale $\Lambda$ range for the ZEUS analysis~\cite{ZEUSCI} 
between 1.7 and 5 TeV. In contrast to the more stringent 
limits obtained by LEP experiments, these bounds do not rely 
on the flavor symmetry hypothesis.
Moreover some models have been considered by ZEUS only.

The CI analysis has been interpreted by the H1 Collaboration
in terms of quark radius, applying a multiplicative form
factor to the SM $d \sigma / d Q^2$. This leads to an upper
limit on the quark radius of $1.9 \times 10^{-16}$ cm~\cite{H1CI}.

\section{Leptoquarks}

Leptoquarks (LQs) are scalar or vector color-triplet bosons,
carrying both lepton ($L$) and baryon ($B$) numbers, which
appear in many extensions of the SM. At HERA, LQs with
fermion number $F=3B+L=0$ ($F=-2$) could be 
resonantly produced via a fusion between the incoming positron
and a quark (antiquark) coming from the proton.

Assuming in particular that a given LQ couples only to known
SM fermions of a given generation,
Buchm\"uller, R\"uckl and Wyler (BRW)~\cite{BUCHMULLER} 
proposed a model in which
the LQs decay exclusively into $e+q$ or $\nu + q$,
with a branching $\beta_e = \beta(LQ \rightarrow eq)$ =
1 or 0.5. Relaxing the hypothesis mentioned above,
other models can be built, where this branching $\beta_e$
is a free parameter.

We will first consider LQs coupling to first generation
fermions only, henceforth called first generation LQs.
The case of Lepton Flavor Violating (LFV) LQs, possessing
couplings both to $eq$ and $\mu q$ or $\tau q$ pairs, will
be addressed in the following paragraph.

 \subsection{First Generation Leptoquarks} 


\begin{figure}
 \begin{center}
\begin{picture}(150,180)(0,0)
  \begin{tabular}{l}
\put(-80,2){\epsfig{file=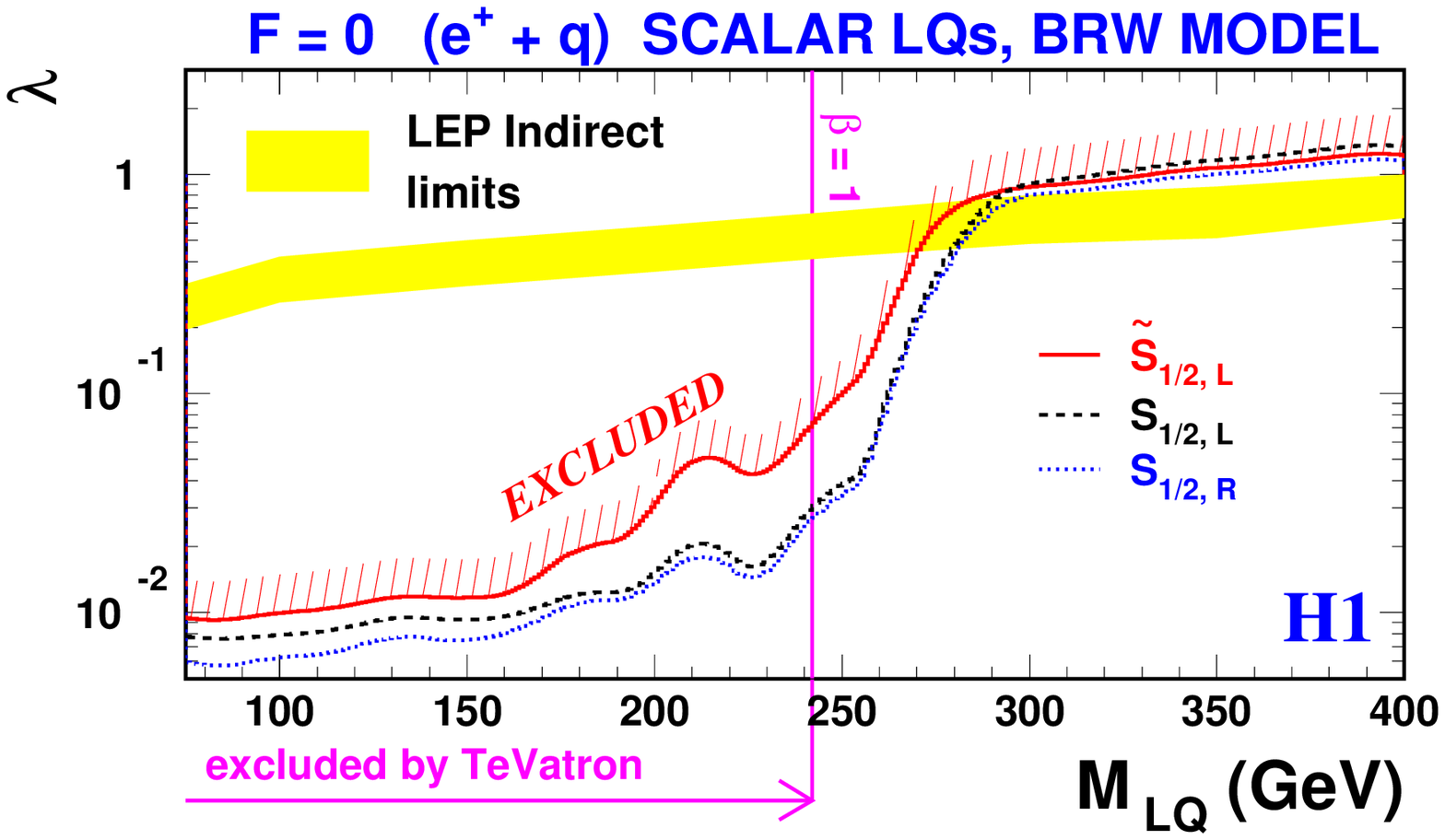,
           width=0.6\textwidth}}
\put(-30,73){{\Large{\mbox{ (a)}}}}
    \\
\put(-40,-2){\epsfig{file=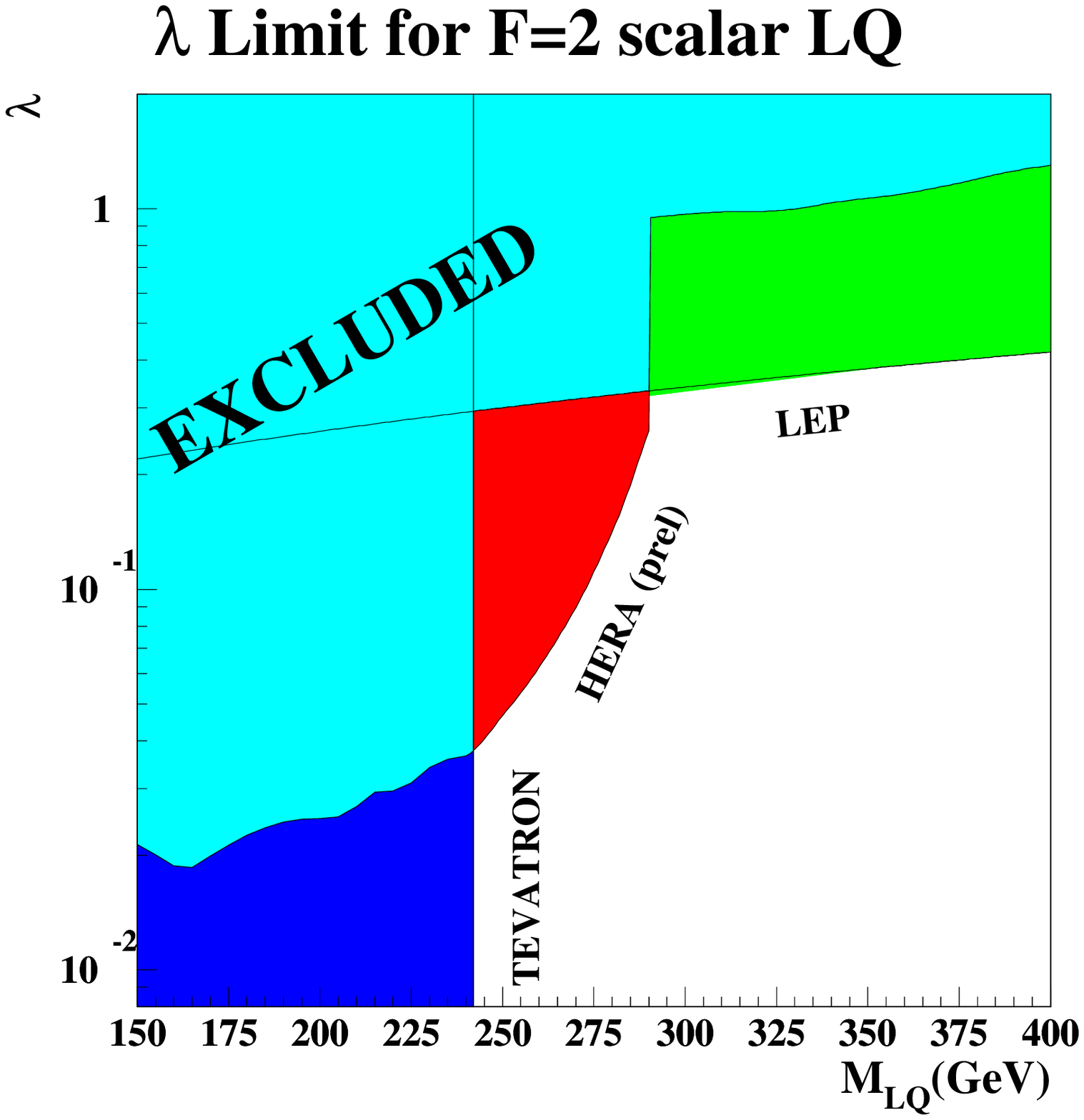,
           width=0.45\textwidth}}
\put(110,40){{\Large{\mbox{ (b)}}}}
   \end{tabular}
\end{picture}
 \end{center}
 \vspace*{6cm}\caption[]{ \label{fig:lqlimits}
   Exlusion limits at $95 \%$ CL on the Yukawa coupling
   $\lambda$ as a function of the LQ mass, for scalar LQs
   with fermion number (a) $F=0$ and
   (b) $\mid F \mid =2$. In (b) a typical LQ example 
   ($S_{1,L}$) is shown. }
\end{figure}

A first generation LQ signal would manifest itself as an excess
of NC DIS-like high $y$ events.
A search for such LQs has been perfomed by H1~\cite{H1LQ} 
and ZEUS~\cite{ZEUSLQ}
in the framework of the BRW model, using the full
$e^+$ dataset.
No significant deviation from the SM has been observed,
apart from a slight excess in H1 data for invariant masses
around 200 GeV, mainly due to events previously reported
in the 94-96 data.
%
The resulting mass-dependent limits on the Yukawa coupling $\lambda$
of the LQ to the $e-q$ pair are shown in Fig.~\ref{fig:lqlimits}a,
for the scalar $F=0$ LQs of the BRW model.
For LQ masses close to or above the kinematic limit, the sensitivity
on the signal is provided by the interference of the LQ processes
with NC DIS and by the effects of finite LQ width.
For a coupling of the electromagnetic strength (i.e.
$\lambda = 0.3$), masses below 275 GeV are ruled out at
$95 \%$ confidence level (CL).
 
The ZEUS Collaboration~\cite{ZEUSLQNC} also analysed 
the recent $e^-$ data accumulated
between 1998 and 1999 at $\sqrt{s} \sim 320$ GeV, corresponding 
to an integrated
luminosity of $\sim 16 {\mbox{pb}}^{-1}$.
No deviation from the SM has been observed and constraints
on $ \mid F \mid =2$ LQs have been set.
The sensitivity achieved is
better than the one provided by the higher statistics $e^+$ data sample,
since with $e^-$ in the initial state the production of
such LQs occurs via a fusion with a quark (instead of a $\bar{q}$).
Typical results obtained by ZEUS 
are shown in 
Fig.~\ref{fig:lqlimits}b.
Scalar $\mid F \mid = 2$ LQs can be excluded up to 290 GeV
for $\lambda = 0.3$.

The case of LQs decaying to $eq$ as well as
to $\nu q$
has also been addressed by both Collaborations~\cite{H1LQ,ZEUSLQCC}.
No signal has been observed in the Charged Current (CC) DIS
sample, and the CC channel has been used to 
better constrain such LQs~\cite{H1LQ}.

\begin{figure}[h]
 \begin{center}
   \hspace*{-0.2cm}\epsfxsize=0.55\textwidth
  \epsffile{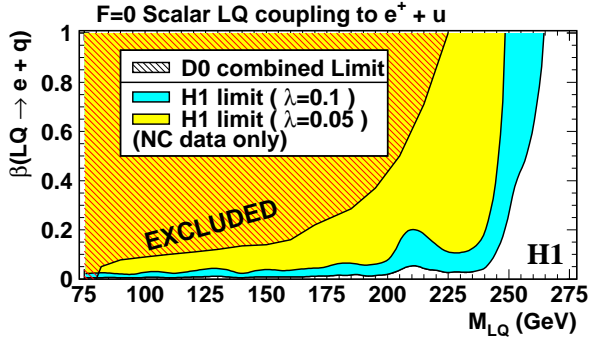}
 \end{center}
 \caption[]{ \label{fig:lqbeta}
  Mass-dependent exclusion limits on the branching
  $\beta(LQ \rightarrow eq)$ for a scalar LQ produced by
  an $e^+ + u$ fusion. Two exclusion domains corresponding
  to $\lambda=0.1$ and $\lambda=0.05$ are represented as shaded
  areas. The $D\emptyset$ limit is also shown as the hatched
  region.}
\end{figure}

Moving away from the BRW model, mass-dependent limits on
the branching $\beta_e$ have been derived by the H1 Collaboration~\cite{H1LQ},
for fixed values of the coupling $\lambda$.
These are shown in Fig.~\ref{fig:lqbeta} for the case of a
generic scalar LQ coupling to $e^+ + u$, for $\lambda=0.1$
and $\lambda=0.05$. The domain covered by H1 extends significantly
beyond the region exluded by the $D\emptyset$ experiment, especially for
low values of $\beta_e$.

 \subsection{Lepton Flavor Violating Leptoquarks}

LQs coupling to $eq$ and $\mu q$ pairs have been considered
by H1~\cite{H1LQ} and ZEUS~\cite{ZEUSLFV}. 
No $e^+ + p \rightarrow \mu + jet$ event compatible with 
LQ kinematics has been found in the $e^+ p$ dataset. 
In particular, the
$e^+ + p \rightarrow \mu + X$ events reported by H1 in~\cite{H1MUEV}
fail significantly the kinematic constraints for LQ-induced
$eq \rightarrow \mu q'$ processes.

Upper bounds on the product $\lambda \times \sqrt{\beta_{\mu}}$,
where $\beta_{\mu}$ denotes the branching 
$\beta(LQ \rightarrow \mu + q)$, have been obtained by ZEUS as
a function of the LQ mass. Scalar (vector) LQs decaying only 
into $eq$ and $\mu q$ 
are excluded up to 278 (285) GeV when
the couplings at the vertices $LQ-e-q$ and $LQ- \mu - q$ are
both of the electromagnetic strength.

The H1 Collaboration~\cite{H1LQ} also performed a search for $\tau + jet$ events,
followed by a hadronic decay of the $\tau$. No event compatible
with LQ kinematics has been observed and constraints on LQs
coupling to $eq$ and $\tau q$ have been derived, showing the
important discovery potential provided by HERA for LQs decaying
with a small branching into $eq$ and a high branching into
$\tau q$.

The case of very high mass ($M_{LQ} \gg \sqrt{s}$) 
LFV LQs was also addressed by
H1 in~\cite{H1LQ}. For both $e \leftrightarrow \mu$
and $e \leftrightarrow \tau$ transitions, direct constraints
on such LQs obtained by H1 were compared to the most stringent
indirect bounds. 
For some LQ types and some $e q_i \leftrightarrow l_j q_k$
reactions H1 limits
extend significantly beyond the reach of other experiments~\cite{H1LQ}.

\section{Excited Fermions}

Excited fermions ($f^*$) would be a clear evidence for fermion
substructure. 
Singly produced $f^*$ (via the $t$-channel exchange
of a gauge boson) have been searched for at HERA, and
results have been interpreted in the framework of the phenomenological
model proposed in~\cite{HAGIWARA}.

\begin{figure}[ht]
 \vspace*{-0.cm}\begin{center}
\begin{picture}(150,200)(0,0)
  \begin{tabular}{l}
\put(-30,2){\epsfig{file=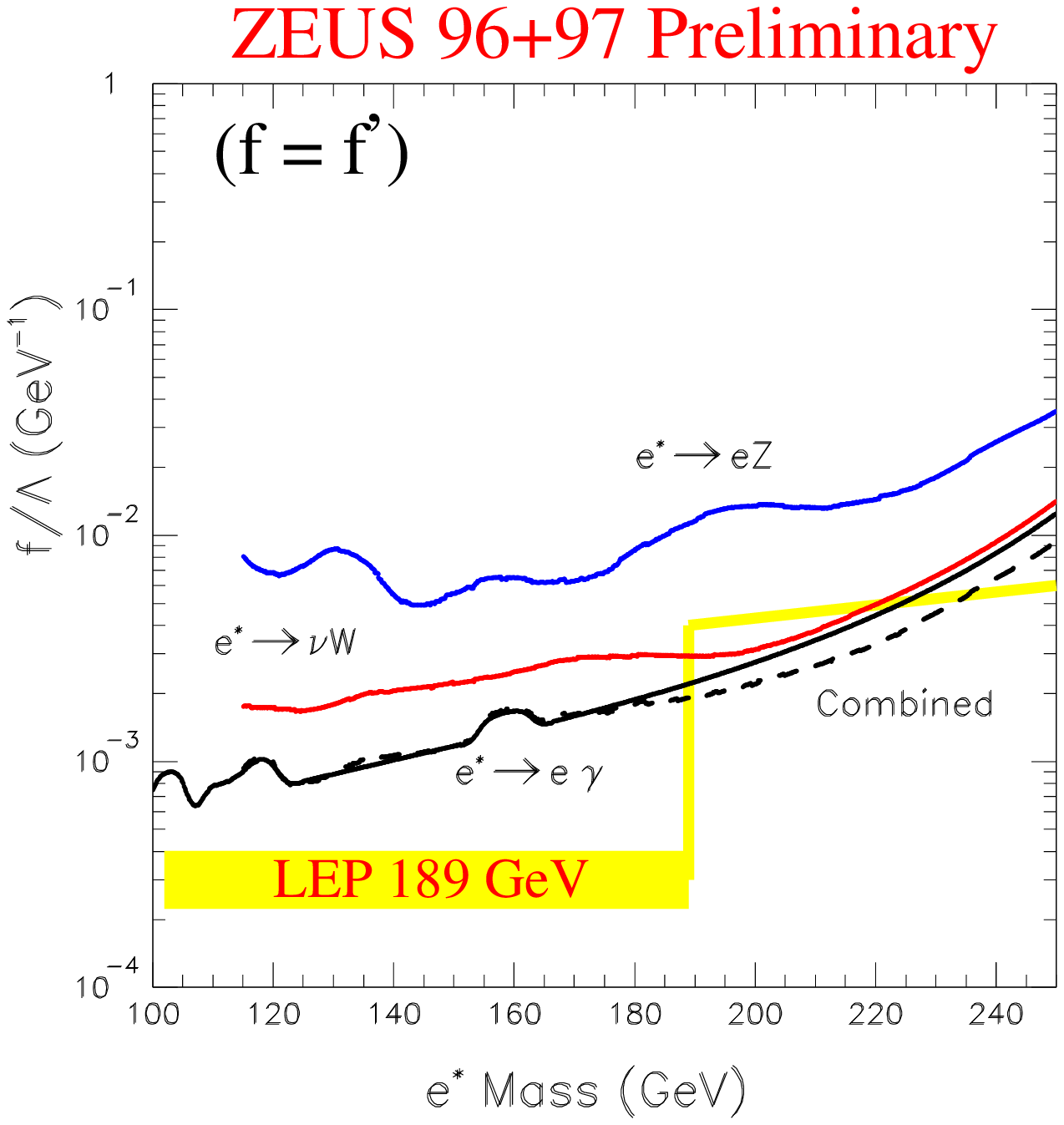,
           width=0.48\textwidth}}
\put(120,170){{\Large{\mbox{ (a)}}}}
    \\
\put(-30,-2){\epsfig{file=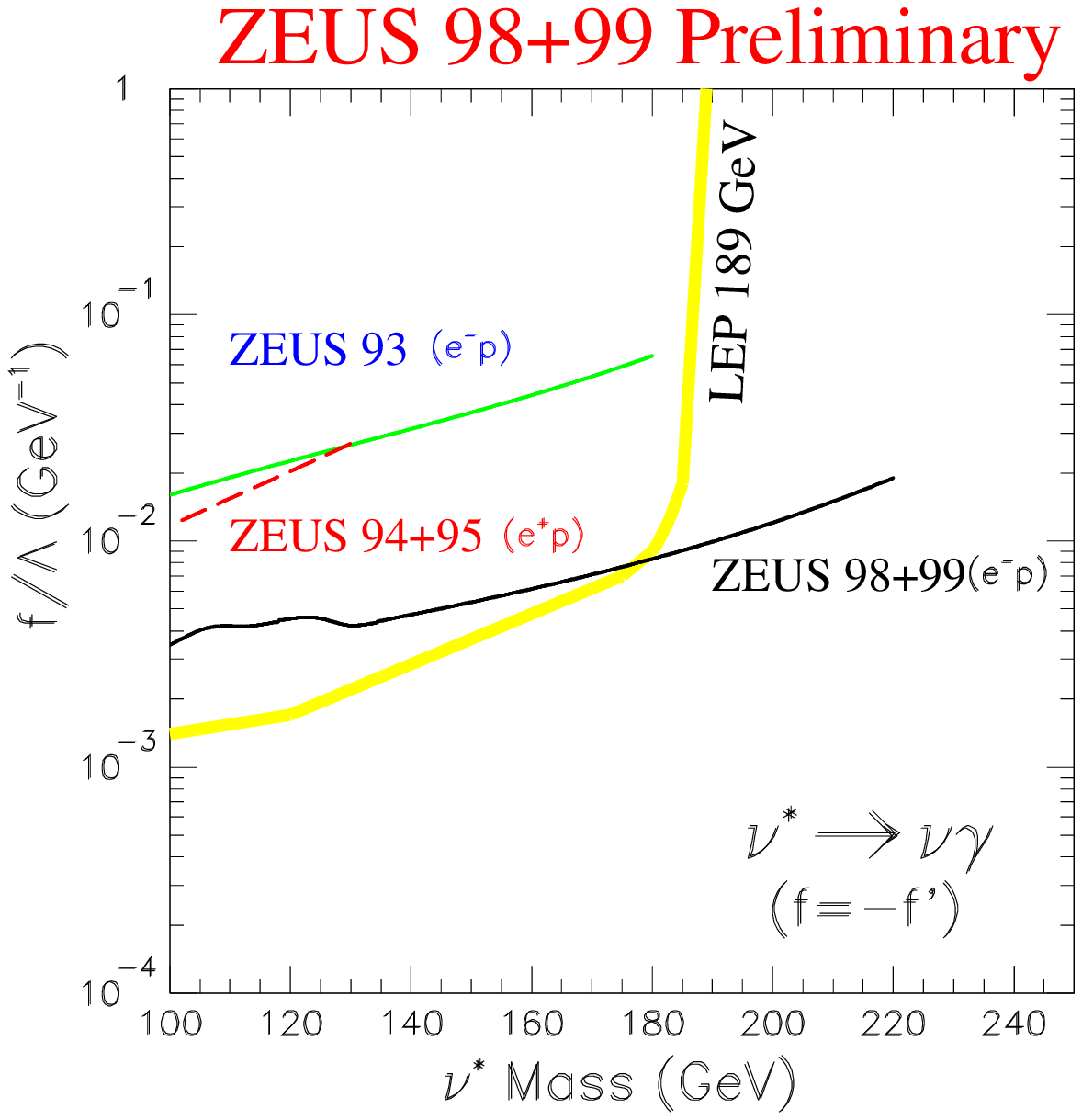,
           width=0.48\textwidth}}
\put(5,45){{\Large{\mbox{ (b)}}}}
   \end{tabular}
\end{picture}
 \end{center}
 \vspace*{7.4cm}\caption[]{ \label{fig:fstar}
   Upper limit at $95 \%$ CL on $f / \Lambda$ as a function of
   (a) the $e^*$ and (b) the $\nu^*$ mass.
   The dashed curve in (a) represents the limit
   obtained combining $e^* \rightarrow e \gamma$,
   $e^* \rightarrow e Z$ and $e^* \rightarrow \nu W$. }
\end{figure}

$e^*$ have been searched for by ZEUS~\cite{ZEUSFSTAR} and
H1~\cite{H1FSTAR} through 
$e^* \rightarrow f + V$ with $V = \gamma, Z, W$,
the electroweak bosons decaying hadronically
or via 
$W \rightarrow e \nu$, $Z \rightarrow e^+ e^-$,
$Z \rightarrow \nu \bar{\nu}$.
%
%
The $e^+p$ data showed no deviation from the SM in all
analysed channels.
Assuming an equal coupling $f$ of the $e^* - e$ pair to $U(1)$
and $SU(2)$ bosons, upper bounds on $f / \Lambda$ 
where $\Lambda$ denotes the compositeness scale
have been obtained as a function of the $e^*$ mass.
%
%
ZEUS limits shown in Fig.~\ref{fig:fstar}a exclude $e^*$ masses below
$229$ GeV at $95 \%$ CL for $f / \Lambda = 1 / M(e^*)$,
and these bounds extend beyond the domain covered by single $e^*$
production at LEP.

%
%
ZEUS~\cite{ZEUSFSTAR} also searched for $\nu^* \rightarrow \nu \gamma$
using
$e^- p$ data, which provide a higher sensitivity than the
$e^+ p$ dataset due to a higher $\nu^*$ production cross-section.
$\nu^*$ lighter
than 161 GeV are excluded for 
$f / \Lambda = 1 / M(\nu^*)$, as can be seen in Fig.~\ref{fig:fstar}b.
 
%
Searches for $q^*$ were reported in~\cite{H1FSTAR} and $q^*$
constraints have been set for a vanishing coupling $f_s$
of the $q^*$ to the gluon, complementary to the stringent bounds
obtained at the TeVatron where $q^*$ production requires $f_s \ne 0$.

%

\section{Conclusions and Outlook}

HERA's discovery potential for
new physics
has been pointed out. 
Although the available data show a general good agreement
with the Standard Model, some observations require clarification.
The five-fold increase of the luminosity as well as
the $e$-beam polarization expected from year 2001
will thus allow exciting searches to be carried out. 



\end{document}